\documentclass[a4paper,11pt]{article}
\usepackage[utf8]{inputenc}
\usepackage{color}
\usepackage{amssymb}
\usepackage{amsmath}
\usepackage{comment}
\usepackage{graphicx}
\usepackage{epstopdf}
\usepackage{mathtools}
\usepackage{ragged2e}
\usepackage{hyperref}
\hypersetup{
	colorlinks=true,
	linkcolor=blue,
	filecolor=cyan, 
	urlcolor=red,
	citecolor=red,
} 
\usepackage[titletoc,toc,title]{appendix}
\numberwithin{equation}{section}
\usepackage{cleveref}
\usepackage[margin=2.5cm]{geometry}
\usepackage{cite}
\usepackage{epsfig}
\usepackage{float}
\usepackage{cancel}
\usepackage{braket}
\usepackage{physics}
\usepackage{amsfonts}
\usepackage{enumitem}
\usepackage[font={footnotesize,it}]{caption}
\usepackage{authblk}

\begin{document}
\begin{titlepage}
\title{Reflected entropy in Galilean conformal field theories and flat holography}

\date{}

\author[1]{Jaydeep Kumar Basak\thanks{\noindent E-mail:~ jaydeep@iitk.ac.in}}

\author[1]{Himanshu Chourasiya\thanks{\noindent E-mail:~ chim@iitk.ac.in}}

\author[1]{Vinayak Raj\thanks{\noindent E-mail:~ vraj@iitk.ac.in}}

\author[1]{Gautam Sengupta\thanks{\noindent E-mail:~ sengupta@iitk.ac.in}}

\affil[1]{
Department of Physics\\
			
Indian Institute of Technology\\ 
			
Kanpur, 208016\\ 
			
India }

\maketitle
\begin{abstract}
\noindent
\justify

We obtain the reflected entropy for bipartite states in a class of $(1+1)$-dimensional Galilean conformal field theories  ($GCFT_{1+1}$) through a replica technique. Furthermore we compare our results with the entanglement wedge cross section (EWCS) obtained for the dual (2+1) dimensional asymptotically flat geometries in the context of flat holography. We find that our results are consistent with the duality between the reflected entropy and the bulk EWCS for flat holographic scenarios.

\end{abstract}
\end{titlepage}
\tableofcontents
\pagebreak


\section{Introduction}
\label{sec1}
\justify
	
In the recent past quantum entanglement in extended many body systems has emerged as a central theme in diverse areas of condensed matter physics and quantum gravity and has seen intense research activity leading to deep insights. It is well known in quantum information theory that the entanglement of bipartite pure states may be characterized by the entanglement entropy which is defined as the von Neumann entropy of the reduced density matrix for the subsystem under consideration. Although the computation of entanglement entropy for finite quantum systems is straightforward the reduced density matrix for quantum many body systems involve infinite number of eigenvalues making it computationally intractable. Interestingly, for conformally invariant $(1+1)$-dimensional quantum field theories ($CFT_{1+1}$), the authors in \cite{Calabrese:2004eu, Calabrese:2005in, Calabrese:2009qy} developed a {\it replica technique} to compute the entanglement entropy.

The characterization of the entanglement for bipartite mixed states in quantum information theory is however a complex issue as the entanglement entropy for such mixed states receives contributions from irrelevant correlations and hence fails to be a viable measure. In this context several computable mixed state correlation and entanglement measures like the entanglement negativity \cite{Vidal:2002zz, Plenio:2005cwa}, the odd entanglement entropy \cite{Tamaoka:2018ned} and the entanglement of purification \cite{Takayanagi:2017knl, Nguyen:2017yqw} and balanced partial entanglement \cite{Wen:2021qgx} have been proposed in the literature\footnote{In quantum information theory many mixed state entanglement measures had been proposed but most of them were difficult to compute as they involved optimization over the local operations and classical communication (LOCC) protocols.}. In the recent past the authors in \cite{Dutta:2019gen} proposed another novel computable correlation measure for mixed state entanglement termed as the {\it reflected entropy}. This is defined as the entanglement entropy of the canonically purified state obtained from the mixed state under consideration. Utilizing an appropriate replica technique the reflected entropy for various bipartite states in $CFT_{1+1}$s was computed in \cite{Dutta:2019gen}. Recently, the authors in \cite{Akers:2021pvd} further explored the reflected entropy in the context of random tensor networks. Furthermore, following the gravitational path integral techniques developed in \cite{Lewkowycz:2013nqa}, they also established a duality between the reflected entropy in holographic $CFT$s and the minimal entanglement wedge cross section (EWCS) for the corresponding dual bulk AdS geometries. We note here that the EWCS has also been proposed as the holographic dual of other measures such as the entanglement of purification \cite{Takayanagi:2017knl}, the balanced partial entanglement \cite{Wen:2021qgx} and the entanglement negativity \cite{Kudler-Flam:2018qjo, Kusuki:2019zsp, KumarBasak:2020eia}\footnote {Note that in a recent communication the authors in \cite{Hayden:2021gno} have introduced a quantity termed as the Markov gap involving the number of non-trivial boundaries of the EWCS. This indicates that the proposed duality in \cite{Kudler-Flam:2018qjo, Kusuki:2019zsp, KumarBasak:2020eia} between the entanglement negativity and the bulk EWCS also involves the Markov gap described in \cite{Hayden:2021gno}.}.

In a different context a class of $(1+1)$-dimensional conformal field theories with Galilean conformal symmetry was obtained in \cite{Bagchi:2009ca, Bagchi:2009my, Bagchi:2009pe} through a parametric \.In\"on\"u-Wigner contraction of the relativistic conformal algebra for $CFT_{1+1}$s. The entanglement entropy for bipartite states in such Galilean conformal field theories ($GCFT_{1+1}$s) was obtained through a replica technique in \cite{Bagchi:2014iea}. In the context of flat space holography \cite{Bagchi:2012cy, Bagchi:2010zz}, the holographic characterization of the entanglement entropy was provided in \cite{Basu:2015evh, Jiang:2017ecm, Hijano:2017eii, Godet:2019wje}. As discussed earlier the entanglement entropy was a valid measure for the entanglement of pure states only which naturally leads to the issue of the characterization of mixed state entanglement in $GCFT_{1+1}$s. In this context, the entanglement negativity for bipartite pure and mixed states in  $GCFT_{1+1}$s  was obtained in \cite{Malvimat:2018izs} employing a replica technique. Subsequently the authors in \cite{Basu:2021axf} proposed a holographic entanglement negativity construction in the framework of flat holography for such bipartite states in $GCFT_{1+1}$s dual to asymptotically flat bulk geometries. Their construction involved the algebraic sums of the areas of codimension-2 extremal surfaces homologous to certain combinations of intervals relevant to the bipartite state configuration in the dual $GCFT_{1+1}$ under consideration which was earlier established in \cite{Chaturvedi:2016rcn, Jain:2017aqk, Malvimat:2018txq} in the context of the $AdS_3/CFT_2$ scenario. Furthermore a novel geometric construction for the bulk EWCS corresponding to bipartite mixed state configurations in the dual $GCFT_{1+1}$s was developed in \cite{Basu:2021awn} for flat space holography\footnote{See \cite{BabaeiVelni:2019pkw, Khoeini-Moghaddam:2020ymm} for the study of entanglement structure in non-relativistic hyperscaling violating theories.}. Very recently, the authors in \cite{Camargo:2022mme} investigated the balanced partial entanglement (BPE) for bipartite mixed states in $GCFT_{1+1}$s and compared their result with the EWCS to verify the duality between the BPE and the EWCS \cite{Wen:2021qgx}.

The above developments bring into sharp focus the issue of the other mixed state correlation measure of the reflected entropy for bipartite states in $GCFT_{1+1}$s and its characterization through the EWCS for the dual bulk asymptotically flat geometries in the context of flat holography. We address this extremely interesting issue in the present article and establish a replica technique for the reflected entropy of bipartite pure and mixed state configurations in $GCFT_{1+1}$s and compare our results with the EWCS computed in \cite{Basu:2021awn} in the context of flat holography. In particular we compute the reflected entropy for bipartite states involving a single, two adjacent and two disjoint intervals in $GCFT_{1+1}$s at zero and finite temperature and for finite sized systems. For the bipartite states involving two disjoint intervals we develop a geometric monodromy analysis first described in \cite{Hijano:2018nhq} to obtain the structure of the dominant Galilean conformal block for the four point twist field correlator required for the reflected entropy of the above mixed state configuration. We find consistent matching of our field theory replica technique results for all the pure and mixed state configurations with the corresponding bulk EWCS in the dual asymptotically flat geometries described in \cite{Basu:2021awn}.

The rest of the article is organized as follows. In \cref{sec2}, we briefly review the reflected entropy in the context of $CFT_{1+1}$s. Subsequently, in \cref{sec3}, following a brief review of $GCFT_{1+1}$s, we obtain the reflected entropy for various bipartite pure and mixed state configurations in such $GCFT_{1+1}$s through a suitable replica technique and compare with the bulk EWCS as mentioned above. In \cref{sec4}, we present a summary of our work and our conclusions. Additionally in Appendix \ref{appendix_A}, we illustrate that the reflected entropy for subsystems in $GCFT_{1+1}$s may also be obtained through a specific non-relativistic limit of the corresponding $CFT_{1+1}$ results.


\section{Review of the reflected entropy}\label{sec2}

\subsection{Reflected entropy}

We begin with a brief review of the reflected entropy in the context of quantum information theory as described in \cite{Dutta:2019gen}. To this end, consider a bipartite quantum system $A \cup B$ in the mixed state $\rho_{AB}$. Its canonical purification $\ket{\sqrt{\rho_{AB}}}$ in a Hilbert space $\mathcal{H}_{A} \otimes \mathcal{H}_{B} \otimes \mathcal{H}_{A^{*}}\otimes \mathcal{H}_{B^{*}}$ involves the CPT conjugate copies $A^*$ and $B^*$ of the subsystems $A$ and $B$ respectively. The reflected entropy $S_{R}(A:B)$ for this bipartite mixed state comprised of the subsystems $A$ and $B$ is defined as the von Neumann entropy of the reduced density matrix $\rho_{AA^{*}}$ as follows 
\begin{equation}\label{def.}
S_{R}(A:B) \equiv S_{vN} (\rho_{AA^{*}})_{\sqrt{\rho_{AB}}}.
\end{equation}
The reduced density matrix $\rho_{AA^{*}}$ is given as
\begin{equation}
\rho_{AA^{*}}= \mathrm{Tr}_{\mathcal{H}_{B} \otimes \mathcal{H}_{B^{*}}} \ket{\sqrt{\rho_{AB}}} \bra{\sqrt{\rho_{AB}}},
\end{equation}
where the degrees of freedom corresponding to the subsystems $B$ and $B^*$ are being traced out.

\subsection{Reflected entropy in $CFT_{1+1}$} \label{sec:SR-review}

\begin{figure}[H]
	\centering
	\includegraphics[scale=.6]{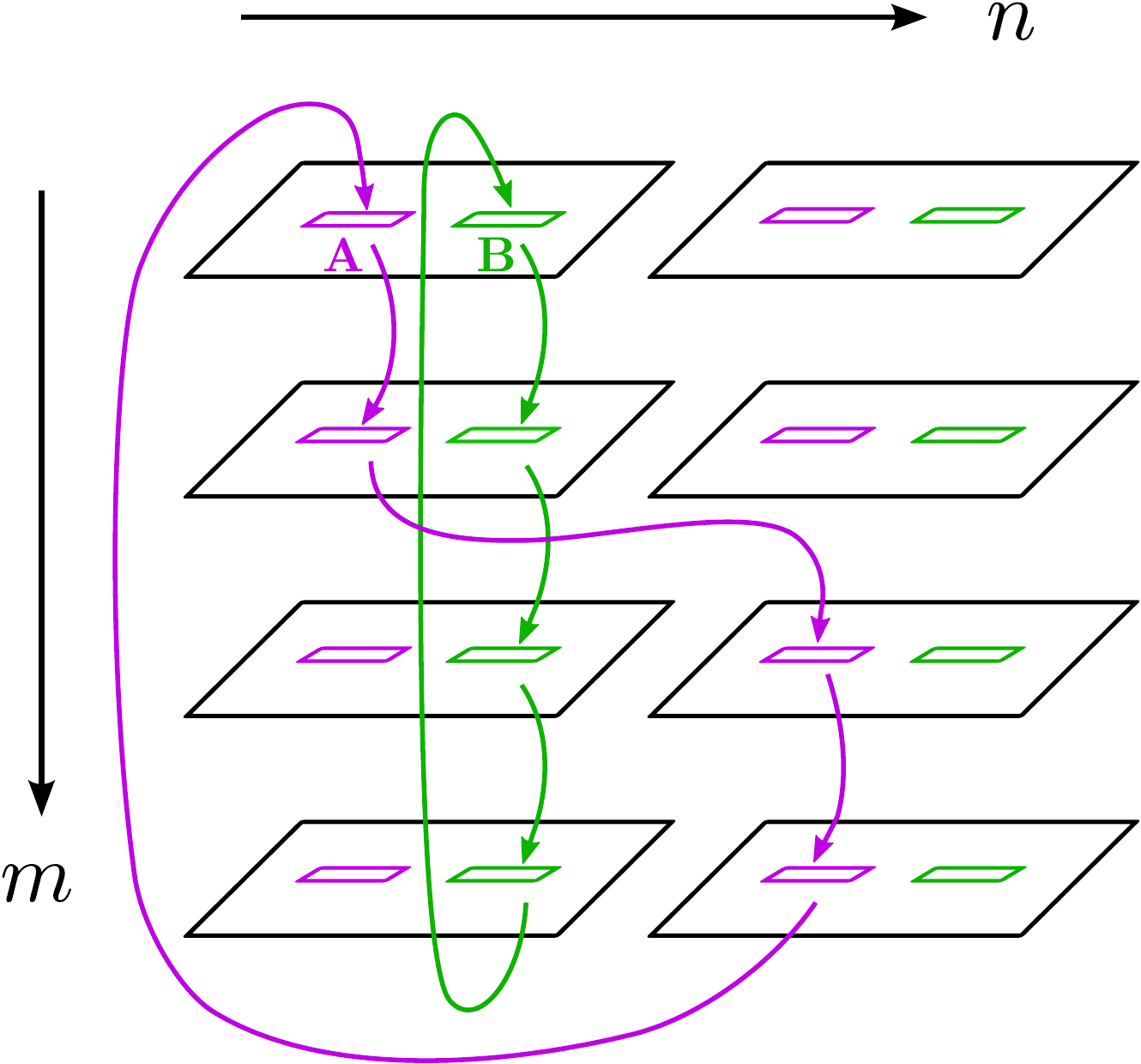}
	\caption{Structure of the replica manifold for the R\'enyi reflected entropy between subsystems $A$ and $B$ in the state $\ket{\psi_m}$. The sewing of the individual replicas along the subsystems $A$ and $B$ are denoted by magenta and green arrows corresponding to the twist fields $\sigma_{g^{}_A}$ and $\sigma_{g^{}_B}$, respectively. Figure modified from \cite{Chandrasekaran:2020qtn}.}
	\label{fig:reflected-replica-simple}
\end{figure}

Interestingly, the authors in \cite{Dutta:2019gen} developed a suitable replica technique which could be utilized to compute the reflected entropy for a bipartite mixed state described by subsystems $A$ and $B$ in arbitrary conformal field theories. One starts with a state $|\rho^{m/2}_{AB}\rangle \equiv \ket{\psi_{m}}$ on a manifold which is constructed by $m$-replication\footnote{See \cite{Dutta:2019gen, Jeong:2019xdr} for details about the replica structure of $|\rho^{m/2}_{AB}\rangle$.} of the original manifold where the subsystems $A$ and $B$ are defined, with $m \in 2\mathbb{Z^{+}}$. The reduced density matrix for this state is given as
\begin{equation}
\rho^{(m)}_{AA^{*}}= \text{Tr}_{\mathcal{H}_{B} \otimes \mathcal{H}_{B^{*}}} \ket{\psi_{m}} \bra{\psi_{m}}.
\end{equation}
The R\'enyi reflected entropy may now be obtained through the R\'enyi entropy $S_n(\rho^{(m)}_{AA^{*}})_{\psi_m}$ which involves another $n$-replication resulting finally in an $nm$-sheeted replica manifold as shown in \cref{fig:reflected-replica-simple}. The reflected entropy for such bipartite states may finally be obtained in the replica limit\footnote{The two replica limits $n \to 1$ and $m \to 1$ are non-commuting as discussed in \cite{Kusuki:2019evw, Akers:2021pvd, Akers:2022max}. In this article, we compute the reflected entropy by first taking $n \to 1$ and subsequently $m \to 1$ as suggested in \cite{Kusuki:2019evw, Akers:2021pvd}.} $n \to 1$ and $m \to 1$ as
\begin{equation}\label{SR-def}
S_{R}(A:B)=\lim_{n,m \to1} S_{n}(AA^{*})_{\psi_{m}}.
\end{equation}

For conformal field theories in $(1+1)$-dimensions ($CFT_{1+1}$s), the reflected entropy may now be computed by the utilization of this replica technique. In this context, the R\'enyi reflected entropy may be obtained in terms of the partition function $Z_{n,m}$ on the $nm$-sheeted replica manifold. This partition function can subsequently be expressed in terms of the correlation function of the twist operators $\sigma_{g_{A}}$ and $\sigma_{g_{B}}$ inserted at the end points of the intervals $A \equiv [z_1, z_2]$ and $B \equiv [z_3, z_4]$ to obtain the R\'enyi reflected entropy as follows \cite{Dutta:2019gen}
\begin{equation}
S_{n}(AA^{*})_{\psi_{m}}=\frac{1}{1-n}\log \frac{Z_{n,m}}{(Z_{1,m})^n}=\frac{1}{1-n} \log \frac{\left<\sigma_{g_{A}}(z_{1})\sigma_{g_{A}^{-1}}(z_{2})\sigma_{g_{B}}(z_{3})\sigma_{g_{B}^{-1}}(z_{4})\right>_{CFT^{\otimes mn}}}{\left(\left<\sigma_{g_{m}}(z_{1})\sigma_{g_{m}^{-1}}(z_{2})\sigma_{g_{m}}(z_{3})\sigma_{g_{m}^{-1}}(z_{4})\right>_{CFT^{\otimes m}}\right)^n} \, .
\end{equation}
In the denominator of the above equation the partition function $Z_{1,m}$ on the $m$-sheeted replica manifold arises from the normalization of the state $|\rho_{AB}^{m/2}\rangle$ and $\sigma_{g_m}$ are the twist fields at the endpoints of the intervals in this $m$-sheeted replica manifold.

In the context of the $AdS/CFT$ duality, following the gravitational path integral techniques developed in \cite{Lewkowycz:2013nqa}, the reflected entropy has also been proved to have a bulk dual in terms of the minimal entanglement wedge cross section (EWCS) as follows \cite{Dutta:2019gen, Akers:2021pvd, Akers:2022max}
\begin{equation}\label{duality}
S_R(A:B) = 2 E_W(A:B),
\end{equation}
where the EWCS $E_W(A:B)$ is defined geometrically as the minimal cross section of the bulk entanglement wedge corresponding to a bipartite quantum state $\rho_{AB}$ \cite{Czech:2012bh, Wall:2012uf}. In this article, we propose to compute the reflected entropy for bipartite states in (1+1)-dimensional Galilean conformal field theories. We will also verify the duality \eqref{duality} by comparing our results with the EWCS obtained in the context of flat space holography in \cite{Basu:2021awn}.



\section{Reflected entropy in Galilean conformal field theories} \label{sec3}

In this section, we first provide a brief review of the (1 + 1)-dimensional Galilean conformal field theories. Subsequent to that we compute the reflected entropy for bipartite states in such $GCFT_{1+1}$s through an appropriate replica technique.

\subsection{Galilean conformal field theories in (1+1)-dimensions} \label{sec:gcft}
In this subsection we briefly recapitulate the salient features of the $(1+1)$-dimensional non-relativistic conformal field theories with Galilean invariance ($GCFT_{1+1}$s) as described in \cite{Bagchi:2009my,Bagchi:2009pe,Bagchi:2009ca}. The conformal algebra for such field theories is described by the Galilean conformal algebra in $(1+1)$-dimensions ($GCA_{1+1}$) which may be obtained from the usual relativistic Virasoro algebra through an In\"on\"u-Wigner contraction described by the rescaling of space and time coordinates as
\begin{equation}\label{Inonu-Wigner}
t\to t,\qquad x\to \epsilon x,
\end{equation}
with $\epsilon\to 0$. This is equivalent to the non-relativistic vanishing velocity limit $v \sim \epsilon$. Any generic Galilean conformal transformation have the following action on the coordinates
\begin{equation}\label{finitetransf}
t\rightarrow f(t)\,,\qquad x\rightarrow f^{\prime}(t)\,x+g(t)\,.
\end{equation}
These can be considered as diffeomorphisms and $t-$dependent shifts respectively. The generators of the $(1+1)$-dimensional GCA in the plane representation are given as follows \cite{Bagchi:2009my}
\begin{equation}\label{GCGT Gen}
L_n=t^{n+1}\partial_t+(n+1)t^nx\partial_x\,, \quad M_n=t^{n+1}\partial_x\,.
\end{equation}
This leads to the lie algebra with different central extensions in each sector as 
\begin{equation}
\begin{aligned}
\left[L_n,L_m\right]&= (m-n)L_{n+m}+\frac{C_L}{12}(n^3-n)\delta_{n+m,0},
\\ [L_n,M_n]&=(m-n)M_{n+m}+\frac{C_M}{12}(n^3-n)\delta_{n+m,0},\\
[M_n,M_m]&=0.
\end{aligned}
\end{equation}
where $C_L$ and $C_M$ are the central charges for the GCA.

Utilizing the Galilean symmetry, one may express the four point correlator for primary fields $V(x,t)$ as \cite{Bagchi:2009pe, Malvimat:2018izs}
\begin{equation}\label{GCFT four point}
\left<\prod_{i=1}^4 V_i(x_i,t_i)\right>=\prod_{1\le i <j \le 4} t_{ij}^{\frac{1}{3}\sum_{k=1}^{4} h_{L,k}-h_{L,i}-h_{L,j}} \,\mathrm{e}^{-\frac{x_{ij}}{t_{ij}}\left (\frac{1}{3} \sum_{k=1}^{4} h_{M,k}-h_{M,i}-h_{M,j}\right)}  \mathcal{G}\left(T,\frac{X}{T}\right),       
\end{equation}
where $x_{ij}=x_i-x_j$, $t_{ij}=t_i-t_j$ and $(h_{L,i},\,h_{M,i})$ are the weights of the primary fields $V_i(x_i,t_i)$. Note that $\mathcal{G}(T,\frac{X}{T})$ is a non universal function which explicitly depend on the specific operator content of the $GCFT_{1+1}$. The cross ratios $T$ and $\frac{X}{T}$ of the $GCFT_{1+1}$, are defined as 
\begin{equation}\label{GCFT cross ratio}
T=\frac{t_{12}t_{34}}{t_{13}t_{24}}\,~, \hspace{1.5cm} \frac{X}{T}=\frac{x_{12}}{t_{12}}+\frac{x_{34}}{t_{34}}-\frac{x_{13}}{t_{13}}-\frac{x_{24}}{t_{24}}\,.
\end{equation}

The entanglement entropy for bipartite states in $GCFT_{1+1}$s could be subsequently computed in \cite{Bagchi:2014iea} through the replica technique involving twist fields. Although an explicit derivation of the Renyi entropy from a path integral over the replica manifold is not available in \cite{Bagchi:2014iea},
such a generalization is plausible. Following \cite{Calabrese:2009qy} the authors in \cite{Bagchi:2014iea} described the GCFT twist fields $\Phi_n$ as primaries under the GCA in the replica limit with conformal weights $h_L^{(n)} = \frac{C_L}{24}\big(n - \frac{1}{n}\big)$ and $h_M^{(n)} = \frac{C_M}{24}\big(n - \frac{1}{n}\big)$ which could be obtained from the GCFT Ward identities. The entanglement entropy for bipartite states in the GCFT was then computed from the two point correlator of these twist fields in \cite{Bagchi:2014iea}. Subsequently such GCFT twist fields were also utilized to obtain the entanglement negativity in \cite{Malvimat:2018izs, Basu:2021axf} and recently for the odd entanglement entropy \cite {Basak:2022gcv} for bipartite states in GCFTs through appropriate replica techniques. The above results also matched with the corresponding bulk holographic computations in the large central charge limit.

Note that, unlike a theory with Lorentz invariance, the choice of a frame affects the observables in $GCFT_{1+1}$s. In order to ascertain this frame dependence Galilean boosted intervals were considered earlier in the literature in relation to the entanglement structure of GCFTs in \cite{Bagchi:2014iea, Basu:2015evh, Jiang:2017ecm, Hijano:2017eii, Malvimat:2018izs, Basu:2021axf, Basak:2022gcv}. In the present article we will consider bipartite states in GCFTs involving such boosted intervals to compute the corresponding reflected entropy. Interestingly such boosted intervals were also considered in \cite{Basu:2015evh, Jiang:2017ecm, Hijano:2017eii} to obtain the entanglement entropy from the bulk dual $(2+1)$-dimensional asymptotically flat geometries in the context of flat holography. Furthermore such boosted intervals were also considered in \cite{Basu:2021axf} for the holographic entanglement negativity and in \cite{Basu:2021awn} for the bulk EWCS in flat holographic scenarios.

\begin{figure}[H]
	\centering
	\includegraphics[scale=.7]{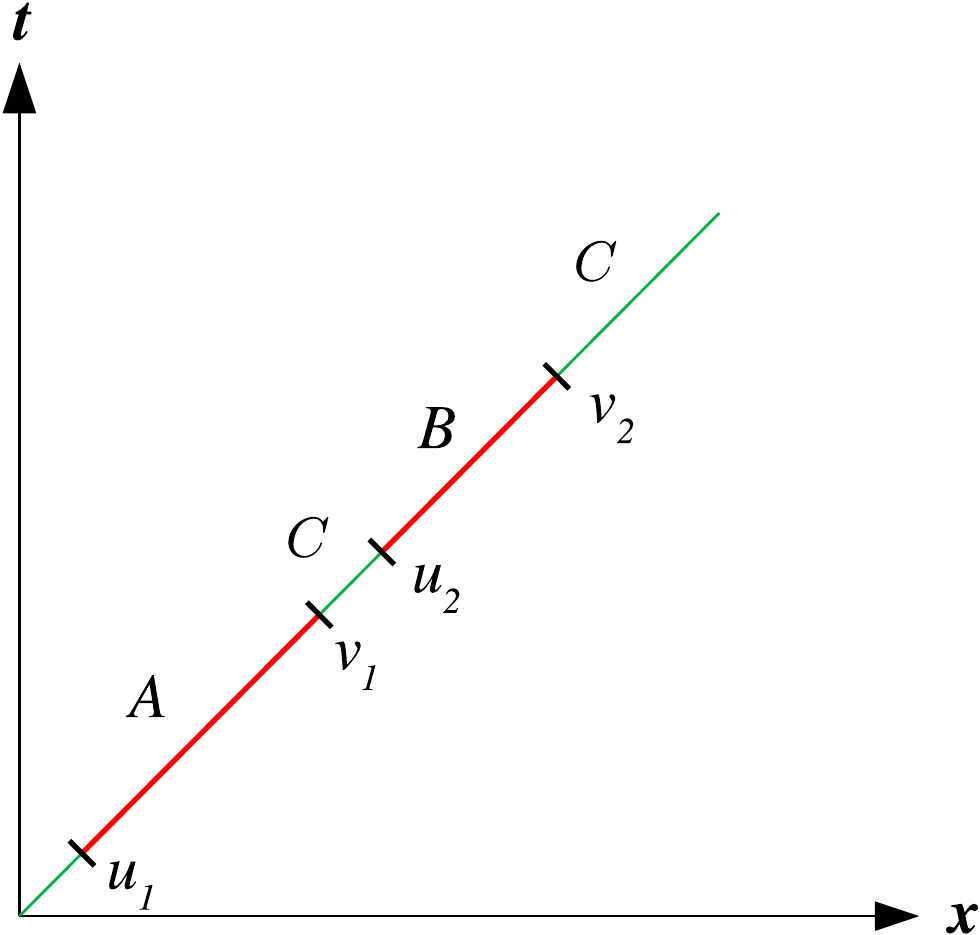}\\
	\caption{Boosted intervals $A$ and $B$ in a  $GCFT_{1+1}$ plane.}
	\label{fig2}
\end{figure}

Similar to the relativistic case, the R\'enyi reflected entropy $S_n(A A^*)_{\psi_m}$ may be computed through a replica technique and may be expressed as a twist field correlator in the $GCFT_{1+1}$ corresponding to the mixed state in question. To illustrate this issue we consider the mixed state configuration of two disjoint boosted intervals $A \equiv [u_1,v_1]$ and $B \equiv [u_2,v_2]$ with $C$ describing the rest of the system as shown in \cref{fig2}. Here $u_1=(x_1,t_1)$, $v_1=(x_2,t_2)$, $u_2=(x_3,t_3)$, $v_2=(x_4,t_4)$ are the end points of the intervals $A$ and $B$ respectively.

Now similar to the case described in \cite{Malvimat:2018izs, Basu:2021axf} in the context of the entanglement negativity, the R\'enyi reflected entropy in the $GCFT_{1+1}$ may be expressed as
\begin{equation}\label{SR-GCFT-def}
S_{n}(AA^{*})_{\psi_{m}}=\frac{1}{1-n}\log \frac{Z_{n,m}}{(Z_{1,m})^n}=\frac{1}{1-n} \log \frac{\left<\sigma_{g_{A}}(u_{1})\sigma_{g_{A}^{-1}}(v_{1})\sigma_{g_{B}}(u_{2})\sigma_{g_{B}^{-1}}(v_{2})\right>_{GCFT^{\otimes mn}}}{\left(\left<\sigma_{g_{m}}(u_{1})\sigma_{g_{m}^{-1}}(v_{1})\sigma_{g_{m}}(u_{2})\sigma_{g_{m}^{-1}}(v_{2})\right>_{GCFT^{\otimes m}}\right)^n}\, ,
\end{equation}
where the partition function $Z_{n, m}$ in the numerator is defined on the $nm$-sheeted $GCFT^{\otimes mn}$ replica manifold and the partition function $Z_{1, m}$ in the denominator is described on the $m$-replicated manifold $GCFT^{\otimes m}$. The twist operators $\sigma_{g_{A}}$ and $\sigma_{g_B}$ appearing in the above expression are similar to the twist fields $\Phi_n$ involved in the replica technique for the entanglement entropy, but are defined on the $nm$-sheeted replica manifold and have the following weights
\begin{equation}\label{reflected-twist-charges}
h_{L} \equiv h^{A}_{L}=h_{L}^{B}=\frac{n \,C_{L}}{24}\left(m-\frac{1}{m}\right),
\end{equation}
with similar expressions for $h_{M} \equiv h^{A}_{M}=h_{M}^{B}$ involving the central charge $C_M$. Also the conformal weights for the twist operator $\sigma_{g_{m}}$ may be obtained from \cref{reflected-twist-charges} by setting $n=1$. In the following subsections we will now compute the reflected entropy for various bipartite state configurations involving a single, two adjacent and two disjoint intervals in the $GCFT_{1+1}$.

\subsection{Reflected entropy for a single interval}
In this subsection we compute the reflected entropy for bipartite pure and mixed states involving a single interval in $GCFT_{1+1}$s.

\subsubsection{Single interval at zero temperature}
For this case, we consider a bipartite pure state of a single boosted interval $A \equiv   [(x_1,t_1),(x_2,t_2)]$, which may be obtained by taking the limit $u_2 \to v_1$ and $v_2 \to u_1$ in the construction described in \cref{SR-GCFT-def} where the interval $ A \cup B$ now describes the full system with $C$ as a null set. In this limit the denominator of \cref{SR-GCFT-def} becomes the identity operator and the numerator reduces to the following two point twist correlator
\begin{equation}\label{single-two-point-gcft}
\log Z_{n,m}=\left<\sigma_{g_{B}^{-1}g_{A}}(u_1)\sigma_{g_{B}g_{A}^{-1}}(v_1)\right>_{GCFT^{\otimes mn}},
\end{equation}
where the twist operators $\sigma_{g_{B}^{-1}g_{A}}$ and $\sigma_{g_{B}g_{A}^{-1}}$ have the following weights
\begin{equation} \label{weight in GCFT}
h^{AB}_{L}=\frac{2 C_L}{24}\left(n-\frac{1}{n}\right)\, ,\qquad  
h^{AB}_{M}=\frac{2 C_M}{24}\left(n-\frac{1}{n}\right).
\end{equation}
\begin{figure}[H]
	\centering
	\includegraphics[scale=.7]{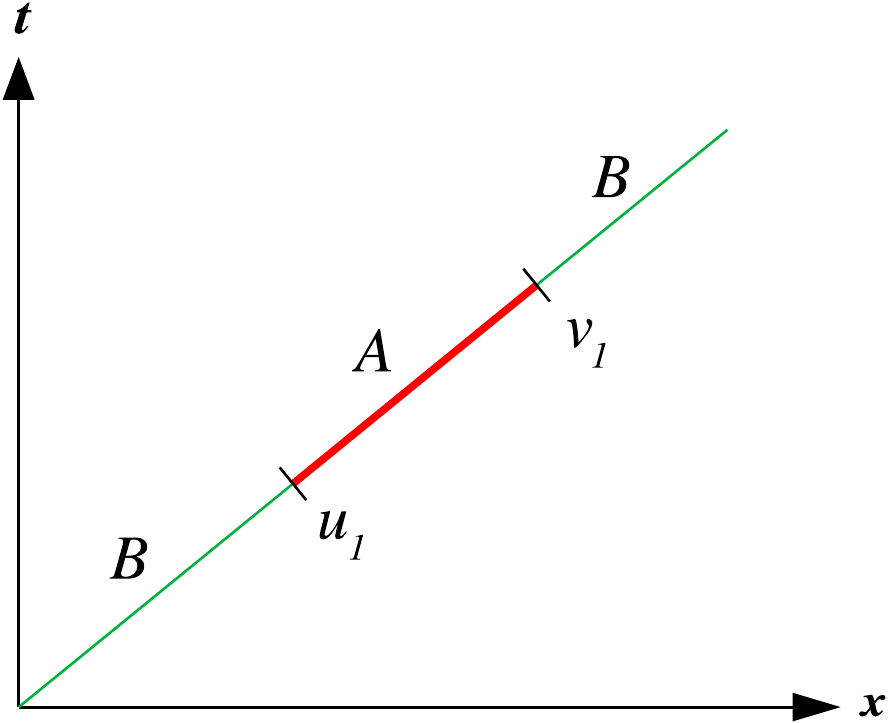}\\
	\caption{A single boosted interval in a  $GCFT_{1+1}$ plane.}
	\label{fig:single}
\end{figure}
By using the usual form of a GCFT two point twist correlator \cite{Bagchi:2009ca} and \cref{SR-def,SR-GCFT-def}, the reflected entropy for the single interval in question may be obtained as
\begin{equation}\label{SR-sing-0}
S_{R} (A:B)= 2 S_A = \frac{C_L}{3} \log \frac{t_{12}}{\epsilon}+\frac{C_M}{3} \frac{x_{12}}{t_{12}}+ \text{const.} \, ,
\end{equation}
where $\epsilon$ is a UV cut-off for the $GCFT_{1+1}$ and the constant arises from the normalization of the two point twist correlator. Note that our field theory result is consistent with the quantum information theory expectation \cite{Dutta:2019gen} that for a pure state, the reflected entropy is equal to twice the entanglement entropy $S_A$ in \cite{Bagchi:2014iea}.

Note here that the $GCFT_{1+1}$ at zero temperature is holographically dual to a bulk $(2+1)$-dimensional asymptotically flat topologically massive gravity (TMG) on a Minkowski space time in the context of flat holography. This is described by a Chern-Simons (CS) term coupled to the usual Einstein-Hilbert action \cite{Hijano:2017eii, Jiang:2017ecm}.
The bulk EWCS for the single interval in this case has been computed in \cite{Basu:2021awn}. It may be observed that our result described by \cref{SR-sing-0} is exactly twice the EWCS computed in \cite{Basu:2021awn}, where the first term involves twice the CS contribution and the second term corresponds to twice the contribution from the Einstein gravity to the bulk EWCS. This demonstrates the consistency of the duality described in \cref{duality} with flat space holography.


\subsubsection{Single interval in a finite size system}
For this case, we consider the bipartite pure state configuration of a single interval in a finite sized $GCFT_{1+1}$ defined on a cylinder with circumference $L$. We may map the $GCFT$ complex plane to this cylinder through the conformal transformation \cite{Malvimat:2018izs, Bagchi:2013qva}
\begin{equation}\label{Trans-size-gcft}
t=e^{\frac{2\pi i\xi}{L}}\,, \qquad x=\frac{2\pi i\rho}{L}\, e^{\frac{2\pi i\xi}{L}}\,,
\end{equation}
where $(x,t)$ are the coordinates on the complex plane and $(\xi, \rho)$ are the coordinates on the cylinder. The $GCFT_{1+1}$ primaries transform under \cref{Trans-size-gcft} as \cite{Malvimat:2018izs, Bagchi:2013qva}
\begin{equation}\label{Trans-primary-size}
\tilde{V_i}(\xi_i,\rho_i)=\left(\frac{L}{2\pi i}\right)^{-h_{L,i}} e^{\frac{2\pi i}{L}(\xi_i h_{L,i}+\rho_i h_{M,i})}V_i(x_i,t_i),
\end{equation}
where $(h_{L,i},\, h_{M,i})$ are the weights of the primaries $V_i$. Using the above transformation in \cref{single-two-point-gcft}, we may obtain the required two point twist correlator on the cylinder as \cite{Malvimat:2018izs}
\begin{equation}\label{two point on cylinder}
\left<\sigma_{g_{B}^{-1}g_{A}}(\xi_1,\rho_1)\sigma_{g_{B}g_{A}^{-1}}(\xi_2,\rho_2)\right>
=\left[\frac{L}{\pi}\sin\left(\frac{\pi \xi_{12}}{L}\right)\right]^{-2h_{L}^{AB}}
\exp\left[-2h_{M}^{AB}\frac{\pi \rho_{12}}{L}\cot\left(\frac{\pi \xi_{12}}{L}\right)\right],
\end{equation}
where $(\xi_i,\rho_i)$ are the endpoints of the interval $A$ on the cylinder. Now using the weights of the twist fields given in \cref{weight in GCFT} we may obtain the reflected entropy for the single interval in question as 
\begin{equation}
S_{R} (A:B) = 2 S_A = \frac{C_L}{3}\log\left(\frac{L}{\pi \epsilon}\sin\frac{\pi \xi_{12}} {L}\right)+ \frac{C_M}{3}\frac{\pi \rho_{12}}{L} \, \cot{\frac{\pi \xi_{12}}{L}}+ \text{const.} \, ,
\end{equation}
where $\epsilon$ is a UV cut-off for the $GCFT_{1+1}$ and the constant is due to the normalization of the corresponding two point twist correlator. Once more it is to be noted that our field theory result matches exactly with twice the entanglement entropy $S_A$ which is consistent with quantum information theory. The corresponding bulk dual in this case is described by 
asymptotically flat TMG on a global Minkowski orbifold. The EWCS for the single interval in this bulk geometry has been computed in \cite{Basu:2021awn}. As earlier it is to be noted that the first term in our field theory computation in the above expression matches with twice the CS contribution and the second term matches with twice the global Minkowski orbifold contribution to the bulk EWCS. This once again describes the consistency of our field theory results with the holographic duality between the reflected entropy and twice the bulk EWCS.


\subsubsection{Single interval at a finite temperature}
In this case we consider the single interval $A \equiv [(-\xi,-\rho),(0,0)]$ in a $GCFT_{1+1}$ at a finite temperature defined on a thermal cylinder whose circumference is equal to the inverse temperature $\beta$. In a very recent article \cite{Basu:2022nds}, it has been shown that in the case of a single interval at a finite temperature in a $CFT_{1+1}$ with an anomaly, a naive computation of the reflected entropy leads to inconsistencies which arises due to the presence of an infinite branch cut. For such a mixed state, the reflected entropy is appropriately obtained through a construction involving two large but finite auxiliary intervals adjacent to the single interval in question on either side. In the present non-relativistic case of a $GCFT_{1+1}$  it is also necessary to consider a similar construction where the single interval in question is sandwiched by two large but finite auxiliary intervals $B_1 \equiv [(-L,-y), (-\xi,-\rho)]$ and $B_2 \equiv [(0,0), (L,y)]$ on either side. The R\'enyi reflected entropy is then obtained with finite auxiliary intervals and finally a bipartite limit $B_1 \cup B_2 \equiv B \to A^c$ is taken to arrive at the original configuration. The reflected entropy for the single interval in question may then be obtained as follows
\begin{equation}\label{single-finite-T(GCFT)}
S_R(A:B)= \lim_{L\to \infty} \lim_{n, m\to 1}\frac{1}{1-n}\log \frac{\left<\sigma_{g_{A}^{-1}}(-L,-y)\sigma_{g_{B}^{-1}g_{A}}(-\xi,-\rho)\sigma_{g_{B}g_{A}^{-1}}(0,0)\sigma_{g_{A}}(L,y)\right>_{GCFT^{\otimes mn}_\beta}}{\left(\left<\sigma_{g_{m}}(-L,-y)\sigma_{g_{m}^{-1}}(L,y)\right>_{GCFT^{\otimes m}_\beta}\right)^n} ,
\end{equation}
where the subscript $\beta$ denotes that the twist correlators are being computed on the thermal cylinder. The four point twist correlator in the numerator of the above equation is given on a GCFT complex plane as follows \cite{Malvimat:2018izs}
\begin{equation} \label{Four pt.(II)(GCFT)}
\begin{aligned}
\Big<\sigma_{g_{A}^{-1}}(x_1, t_1)\sigma_{g_{B}^{-1}g_{A}}(x_2, t_2)\sigma_{g_{B}g_{A}^{-1}}(x_3, t_3)\sigma_{g_{A}}(x_4, t_4)&\Big>_{GCFT^{\otimes mn}}
=\frac{k_{mn}^2}{t_{14}^    
	{2 h_{L}}\,t_{23}^{2 h^{AB}_{L}}}
\frac{\mathcal{F}_{mn}\left(T,\frac{X}{T}\right)}{T^{h^{AB}_{L}
}}\\	
& \times \exp \Bigg [ - 2 h_{M}
\frac{x_{14}}{t_{14}}
- 2 h^{AB}_{M} \frac{x_{23}}{t_{23}} - h^{AB}_{M} \frac{X}{T} \Bigg ],
\end{aligned}
\end{equation}
where $X/T$ and $T$ are the $GCFT_{1+1}$ cross ratios given in \cref{GCFT cross ratio} and the  non-universal function $\mathcal{F}_{mn}$ may be obtained in the limits $T \to 1$ and $T \to 0$ as \cite{Malvimat:2018izs}
\begin{equation}
\mathcal{F}_{mn}(1,0)=1, \hspace{2cm} \mathcal{F}_{mn}\left(0,\frac{X}{T}\right)= C_{mn}.
\end{equation} 
Here $C_{mn}$ is a non-universal constant that depends on the full operator content of the theory.

We may utilize the following conformal map to transform the $GCFT_{1+1}$ plane with coordinates $(x,t)$ to the thermal cylinder with coordinates $(\xi,\rho)$ \cite{Bagchi:2013qva, Malvimat:2018izs}:
\begin{equation}\label{Trans-Temp-gcft}
t=e^{\frac{2\pi\xi}{\beta}}, \qquad x=\frac{2\pi\rho}{\beta}e^{\frac{2\pi\xi}{\beta}}.
\end{equation}
The $GCFT_{1+1}$ primaries transform under the above conformal map as\cite{Bagchi:2013qva, Malvimat:2018izs}
\begin{equation}\label{Trans-primary-Temp}
\tilde{V}_i(\xi_{i},\rho_{i})=\left(\frac{\beta}{2\pi}\right)^{-h_{L,i}} e^{\frac{2\pi}{\beta}(\xi_{i} h_{L,i}+\rho_{i} h_{M,i})}\,V_i(x_i,t_i).
\end{equation}
We may obtain the required four point twist correlator on the cylinder by using \cref{Trans-Temp-gcft,Trans-primary-Temp} in \cref{Four pt.(II)(GCFT)} as
\begin{equation}\label{FINITE}
\begin{aligned}
\Big<\sigma_{g_{A}^{-1}}&(-L,-y)\sigma_{g_{B}^{-1}g_{A}}(-\xi,-\rho)\sigma_{g_{B}g_{A}^{-1}}(0,0)\sigma_{g_{A}}(L,y)\Big>_{GCFT^{\otimes mn}_\beta}\\
&\quad  = \frac{k_{mn}^2}
{T^{h^{AB}_{L}}}\left [\frac{\beta}
{\pi}\sinh\left( \frac{2\pi L}{\beta}\right )\right ]^{-2 h_{L}}
\left [\frac{\beta}{\pi}\sinh \left(\frac{\pi\xi}{\beta}\right)\right ]^{-2 h^{AB}_{L}} \exp\bigg[-\frac{2\pi y}{\beta}\coth\left(\frac{2\pi L}{\beta}\right)2h_{M}\\
&\quad \qquad \qquad \qquad \qquad \qquad \qquad \qquad \quad -\frac{2\pi \rho}{\beta}\coth\left(\frac{\pi\xi}{\beta}\right)h^{AB}_{M}-\frac{X}{T}
h^{AB}_{M}\bigg]\mathcal{F}_{mn}\left(T, \frac{X}{T}\right).
\end{aligned}
\end{equation}
In the bipartite limit, the cross-ratios $X$ and $T$ transformed under \cref{Trans-Temp-gcft} have the following form
\begin{equation}\label{limit}
\lim_{L \to \infty} T= \exp(-\frac{2 \pi \xi}{\beta}), \quad \quad \lim_{L \to \infty} \frac{X}{T}=-\frac{2 \pi \rho}{\beta}.
\end{equation}
Now by substituting \cref{FINITE,limit} in \cref{single-finite-T(GCFT)} and taking the bipartite limit $L \to \infty$ subsequent to the replica limit $n \to 1$ and $m \to 1$, we may obtain the reflected entropy for the single interval $A$ at a finite temperature $\beta$ to be
\begin{equation}\label{result finite T}
S_{R} (A:B)= \frac{C_L}{3} \left[\log\left(\frac{\beta}{\pi \epsilon}\sinh\frac{\pi \xi}{\beta}\right)-\frac{\pi \xi}{\beta}\right]+\frac{C_M}{3} \left[\frac{\pi \rho}{\beta}\coth\frac{\pi \xi}{\beta}-\frac{\pi \rho}{\beta}\right]+ f\left(e^{-\frac{2\pi\xi}{\beta}},-\frac{2\pi\rho}{\beta}\right)+ \ldots,
\end{equation}
where $\epsilon$ is a UV cut off and the non universal function $f (T,X/T)$ is given as \cite{Malvimat:2018izs}
\begin{equation}
f\left(T,\frac{X}{T}\right) \equiv \lim_{n, m \to 1} \ln \left[ \mathcal{F}_{mn} \left(T, \frac{X}{T} \right) \right].
\end{equation}
We observe that it is possible to express the reflected entropy in \cref{result finite T} in a more instructive way as follows
\begin{equation}
S_R(A:B) = 2\left[S_A-S_A^\text{th}\right] + f\left(e^{-\frac{2\pi\xi}{\beta}},-\frac{2\pi\rho}{\beta}\right)+ \ldots,
\end{equation}
where $S_A$ denotes the entanglement entropy of the single interval $A$ \cite{Bagchi:2014iea} and $S_A^\text{th}$ denotes the thermal contribution. This is indicative of the absence of the thermal correlations in the universal part of the reflected entropy. The holographic dual for this case is described by bulk non-rotating flat space cosmologies (FSC) with topologically massive gravity. As earlier we observe that our field theory result in \cref{result finite T} matches with twice the upper bound of the EWCS computed in \cite{Basu:2021awn} apart from an additive constant contained in the non universal function $f (T,X/T)$ which may be extracted through a large central charge analysis of the corresponding conformal block. Here also the first term matches with twice the CS contribution while the second term with twice the FSC contribution to the bulk EWCS consistent with the holographic duality mentioned earlier.


\subsection{Reflected entropy for adjacent intervals}		
We now proceed to the computation of the reflected entropy for the bipartite mixed states of two adjacent intervals in non-relativistic $GCFT_{1+1}$s. In particular we consider two adjacent intervals at zero and a finite temperature, and in a finite sized system.

\subsubsection{Adjacent intervals at zero temperature}		

In this case we consider the bipartite mixed state of two adjacent intervals $A\equiv [u_1,u_2]$, $B \equiv [u_2,v_2]$ in $GCFT_{1+1}$ at  zero temperature. This configuration may be obtained by taking the adjacent limit $v_1 \to u_2$ in the disjoint intervals example considered in subsection \ref{sec:gcft}. In this adjacent limit, the R\'enyi reflected entropy in \cref{SR-GCFT-def} reduces to
\begin{equation} \label{renyi-adj}
S_{n}(AA^{*})_{\psi_{m}}=\frac{1}{1-n} \log \frac{\left<\sigma_{g_{A}}(u_{1})\sigma_{g_{B}g_{A}^{-1}}(u_{2})\sigma_{g_{B}^{-1}}(v_{2})\right>_{GCFT^{\otimes mn}}}{\left(\left<\sigma_{g_{m}}(u_{1})\sigma_{g_{m}^{-1}}(v_{2})\right>_{GCFT^{\otimes m}}\right)^n}.
\end{equation}
\begin{figure}[H]
	\centering
	\includegraphics[scale=.7]{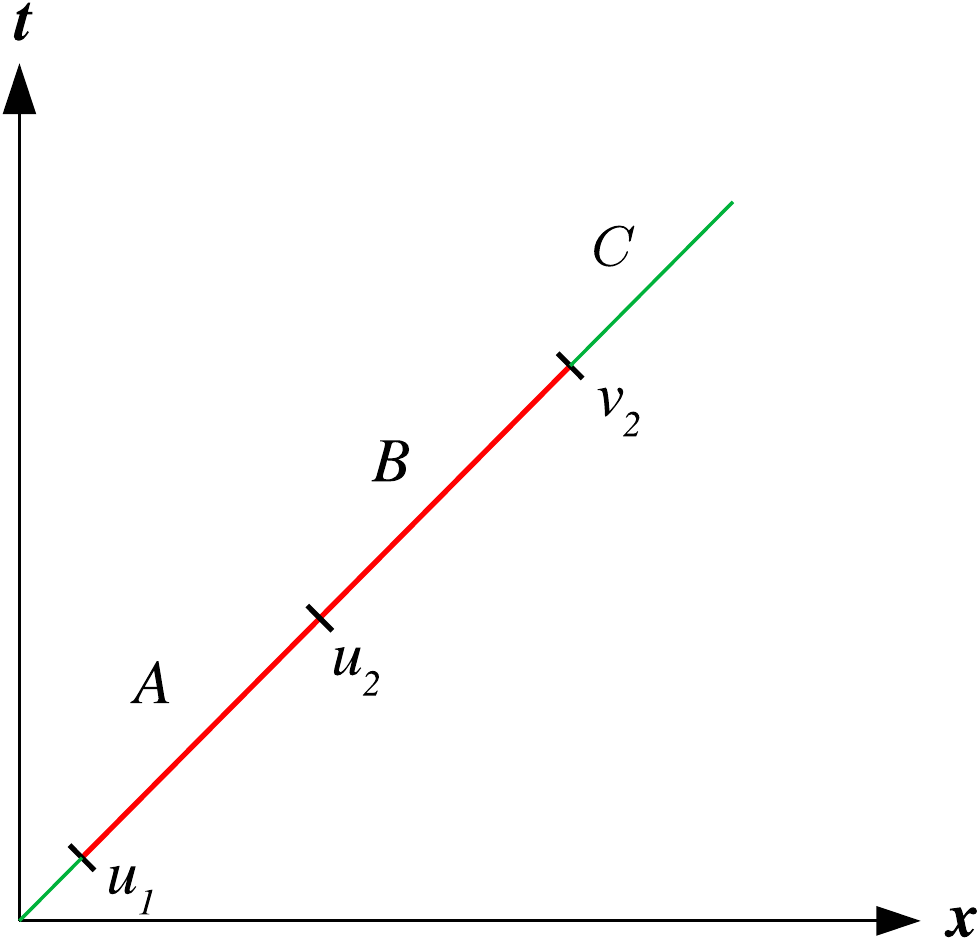}\\
	\caption{Two adjacent intervals in a  $GCFT_{1+1}$ plane.}
	\label{fig3}
\end{figure}
Using the usual structure of $GCFT_{1+1}$ correlation functions \cite{Bagchi:2009ca} and taking the replica limit $n \to 1$ and $m \to 1$ in \cref{renyi-adj}, we may obtain the reflected entropy for the mixed state of two adjacent intervals under consideration as
\begin{equation}
S_{R}(A:B)= \frac{C_L}{6}\log\left(\frac{t_{12}t_{23}}{\epsilon t_{13}}\right) +\frac{C_M}{6}\left(\frac{x_{12}}{t_{12}}+\frac{x_{23}}{t_{23}}-\frac{x_{13}}{t_{13}}\right) + \ldots,
\end{equation}		
where $\epsilon$ is a UV cut-off. As mentioned earlier $GCFT_{1+1}$ at zero temperature is holographically dual to a bulk $(2 + 1)$- dimensional asymptotically flat TMG on a Minkowski space time in the context of flat holography. Again, we observe that in the large central charge limit, our result matches with twice the bulk EWCS in \cite{Basu:2021awn} apart from an additive constant arising from the undetermined OPE coefficient of the three point twist correlator in \cref{renyi-adj}. Here the first term corresponds to twice the CS contribution and the second term involves twice the contribution from the Einstein gravity to the bulk EWCS.


\subsubsection{Adjacent intervals in a finite size system}		

We now proceed to the bipartite mixed state configuration of two adjacent intervals in a finite sized $GCFT_{1+1}$ described on a cylinder of circumference $L$. To this end we consider the following adjacent intervals, $A \equiv [(\xi_1,\rho_1), (\xi_2,\rho_2)]$ and $B \equiv [(\xi_2,\rho_2), (\xi_3,\rho_3)]$. Using the transformation of primaries given in \cref{Trans-primary-size} under the conformal map in \cref{Trans-size-gcft}, we may obtain the required three point twist correlator on the cylinder as follows
\begin{equation}\label{Adj. Trans. Size}
\begin{aligned}
\Big<\sigma_{g_{A}}(\xi_1,\rho_1)\sigma_{g_{B}g_{A}^{-1}}(\xi_2,\rho_2)\sigma_{g_{B}^{-1}}(\xi_3,&\rho_3)\Big>
=\left(\frac{L}{2\pi i}\right)^{-2h_{L}-h_{L}^{AB}} 
\exp\Bigg[\frac{2\pi i}{L}\bigg(\xi_1h_{L}+\xi_2h_{L}^{AB}+\xi_3h_{L}\\
&+\rho_1h_{M}+\rho_2h_{M}^{AB}+\rho_3h_M\bigg)\Bigg]\,
\left<\sigma_{g_{A}}(u_{1})\sigma_{g_{B}g_{A}^{-1}}(u_{2})\sigma_{g_{B}^{-1}}(v_{2})\right>,
\end{aligned}
\end{equation}	
where $\big(h_{L}, h_{M}\big), \, \big(h_{L}^{AB}, h_{M}^{AB}\big)$ are the weights of the $GCFT_{1+1}$ twist fields $\sigma_{g_{A}}$ and $\sigma_{g_{B}g_{A}^{-1}}$ respectively and the three point twist correlator on the right-hand-side is defined on the $GCFT$ plane. We may now obtain the reflected entropy for the bipartite mixed state of two adjacent intervals under consideration using \cref{Adj. Trans. Size} in \cref{SR-def,SR-GCFT-def} as follows
\begin{equation}\label{adj-L}
\begin{split}
S_{R}(A:B) = \frac{C_L}{6} \log\left[\frac{L}{\pi \epsilon} \frac{\sin\left(\frac{\pi \xi_{12}}{L}\right)\sin\left(\frac{\pi \xi_{23}}{L}\right)}{\sin\left(\frac{\pi \xi_{13}}{L}\right)}\right] +\frac{C_M}{6 }\Bigg[\frac{\pi \rho_{12}}{L}\cot&\left(\frac{\pi \xi_{12}}{L}\right)+\frac{\pi \rho_{23}}{L} \cot\left(\frac{\pi \xi_{23}}{L}\right)\\
&-\frac{\pi \rho_{13}}{L} \cot\left(\frac{\pi \xi_{13}}{L}\right)\Bigg]+\ldots.
\end{split}
\end{equation}
Note that the bulk dual for this case is described by asymptotically flat TMG on a global Minkowski orbifold. Again, we observe that in the large central charge limit, the first term of the above expression involves twice the CS contribution while the second term corresponds to twice the global Minkowski orbifold contribution to the EWCS \cite{Basu:2021awn}, modulo an additive constant arising from the undetermined OPE coefficient of the three point function in \cref{Adj. Trans. Size}.


\subsubsection{Adjacent intervals at a finite temperature} \label{sec:adj-T-GCFT}		

For this case we consider the mixed state configuration of two adjacent interval described by $A \equiv [(\xi_1,\rho_1), (\xi_2,\rho_2)]$ and $B \equiv [(\xi_2,\rho_2), (\xi_3,\rho_3)]$ at a finite temperature in a $GCFT_{1+1}$ defined on a thermal cylinder with the circumference given by the inverse temperature $\beta$. We may again utilize the transformation of the primaries in \cref{Trans-primary-Temp} under the action of the conformal map \eqref{Trans-Temp-gcft} from the $GCFT$ plane to the thermal cylinder to obtain the corresponding three point twist correlator as
\begin{equation}\label{ft}
\begin{aligned}
\Big<\sigma_{g_{A}}(\xi_1,\rho_1)\sigma_{g_{B}g_{A}^{-1}}&(\xi_2,\rho_2)\sigma_{g_{B}^{-1}}(\xi_3,\rho_3)\Big>_{GCFT^{\otimes mn}_\beta}
=\left(\frac{\beta}{2\pi }\right)^{-2h_{L}-h_{L}^{AB}} 
\exp\Bigg[\frac{2\pi}{\beta}(\xi_1h_{L}+\xi_2h_{L}^{AB}\\
&+\xi_3h_{L}+\rho_1h_{M}+\rho_2h_{M}^{AB}+\rho_3h_M)\Bigg]\,
\left<\sigma_{g_{A}}(u_{1})\sigma_{g_{B}g_{A}^{-1}}(u_{2})\sigma_{g_{B}^{-1}}(v_{2})\right>_{GCFT^{\otimes mn}}.
\end{aligned}
\end{equation}
Here the subscript $\beta$ denotes that the correlator is being computed on the thermal cylinder and the three point twist correlator on the right-hand-side is defined on the $nm$-replicated $GCFT$ plane. Now using the usual form of the three point correlator in the $GCFT$ plane \cite{Bagchi:2009ca} in the above expression, we may obtain the reflected entropy for two adjacent intervals at a finite temperature as
\begin{equation}\label{adj-T}
\begin{split}
S_R(A:B) = \frac{C_L}{6} \log\Bigg[\frac{\beta}{\pi \epsilon} \frac{\sinh\left(\frac{\pi \xi_{12}}{\beta}\right)\sinh\left(\frac{\pi\xi_{23}}{\beta}\right)}{\sinh\left(\frac{\pi \xi_{13}}{\beta}\right)}\Bigg]   +\frac{C_M}{6 }\Bigg[&\frac{\pi \rho_{12}}{\beta} \coth\left(\frac{\pi \xi_{12}}{\beta}\right) \\ +\frac{\pi \rho_{23}}{\beta}\coth\left(\frac{\pi \xi_{23}}{\beta}\right)
&-\frac{\pi \rho_{13}}{\beta} \coth\left(\frac{\pi \xi_{13}}{\beta}\right)\Bigg]+\ldots,
\end{split}
\end{equation}		
where $\epsilon$ is a UV cut-off of the $GCFT_{1+1}$. As mentioned earlier the holographic dual in this case is described by a bulk non rotating FSC with TMG. Note that in the large central charge limit, the above result matches with twice the EWCS computed in \cite{Basu:2021awn} apart from an additive constant contained in the undetermined OPE coefficient of the three point twist correlator in \cref{ft}. Here also the first term matches with twice the CS contribution whereas the second term corresponds to twice the FSC contribution to the bulk EWCS.


\subsection{Reflected entropy for two disjoint intervals}

Finally in this subsection we focus on the mixed state configuration of two disjoint intervals in $GCFT_{1+1}$s. To this end we consider two disjoint intervals $A \equiv [(x_1, t_1), (x_2, t_2)]$ and $B \equiv [(x_3, t_3), (x_4, t_4)]$. The computation of reflected entropy in this case requires the monodromy analysis of the four point twist correlator in \cref{SR-GCFT-def}. The numerator of \cref{SR-GCFT-def} may be expanded in terms of the Galilean conformal blocks $\mathcal{F}_{\alpha}$ corresponding to the $t$-channel ($T \to 1 $, $X \to 0$) as follows 
\begin{equation}\label{4pointmonodromy}
\begin{aligned}
\Big<\sigma_{g_{A}}(x_{1},t_{1})\sigma_{g_{A}^{-1}}(x_{2},t_{2})\sigma_{g_{B}}(x_{3},t_{3})&\sigma_{g_{B}^{-1}}(x_{4},t_{4})\Big>_{GCFT^{\otimes mn}}\\
& = t_{14}^{-2h_{L}} \, t_{23}^{-2h_{L}} \, \exp\left[-2h_{M} \frac{x_{14}}{t_{14}}-2h_{M} \frac{x_{23}}{t_{23}}\right]\,\sum_{\alpha}\, \mathcal{F}_{\alpha} \left( T, \frac{X}{T} \right).
\end{aligned}
\end{equation}
The conformal blocks $\mathcal{F}_{\alpha}$ are arbitrary functions of the cross ratios and depend on the full operator content of the theory. However in the large central charge limit $C_L, C_M \to \infty$, similar to the relativistic case discussed in \cite{Fitzpatrick:2014vua}, the blocks $\mathcal{F}_{\alpha}$ are expected to have an exponential structure which implies that the dominant contribution to the four point twist correlator arises from the conformal block with the lowest conformal weight. We will extract the structure of these $GCFT$ conformal blocks $\mathcal{F}_{\alpha}$ through an appropriate (geometric) monodromy analysis \cite{Hijano:2018nhq}.

Note that unlike the relativistic $CFT_{1+1}$ case, in $GCFT_{1+1}$ the two components of the energy-momentum tensor are not identical and are given as \cite{Hijano:2018nhq}
\begin{equation}\label{stressTensors}
\mathcal{M}\equiv T_{tx}=\sum_n M_n\,t^{-n-2}\, , \qquad \mathcal{L}\equiv T_{tt}=\sum_n\left[L_n+(n+2)\frac{x}{t}M_n\right]\,t^{-n-2} \, ,
\end{equation}
where $L_n$ and $M_n$ are the GCA generators defined in \cref{GCGT Gen}. For the two components of the energy-momentum tensor $\mathcal{M}$ and $\mathcal{L}$, the Galilean Ward identities are given as \cite{Hijano:2018nhq}
\begin{equation}\label{WardIdentity}
\begin{aligned}
&\left<\mathcal{M}(x,t)V_1(x_1,t_1)\dots V_n(x_n,t_n)\right>=\sum_{i=1}^n \left[\frac{h_{M,i}}{(t-t_i)^2}+\frac{1}{t-t_i}\partial_{x_i}\right]\left< V_1(x_1,t_1)\dots V_n(x_n,t_n)\right>\,,\\
&\left<\mathcal{L}(x,t)V_1(x_1,t_1)\dots V_n(x_n,t_n)\right>=\sum_{i=1}^n \bigg[\frac{h_{L,i}}{(t-t_i)^2}-\frac{1}{t-t_i}\partial_{t_i}+\frac{2h_{M,i}(x-x_i)}{(t-t_i)^3}\\
&\qquad\qquad\qquad\qquad\qquad\qquad\qquad\qquad\quad+\frac{x-x_i}{(t-t_i)^2}\partial_{x_i}\bigg]\left< V_1(x_1,t_1)\dots V_n(x_n,t_n)\right>\,,
\end{aligned}
\end{equation}
where $V_i$s are  $GCFT_{1+1}$ primaries with $(h_{L,i},\, h_{M,i})$ being their corresponding weights.

In the following subsections we will implement separate geometric monodromy analysis \cite{Hijano:2018nhq, Basu:2021axf} for each of the two components $\mathcal{M}$ and $\mathcal{L}$ in the semi classical limit of large central charge to obtain the expression for the conformal blocks $F_{\alpha}$. Subsequently we will utilize these results to obtain the reflected entropy for bipartite mixed states involving the two disjoint intervals in $GCFT_{1+1}$s.


\subsubsection{Monodromy of $\mathcal{M}$} \label{M-monodromy}

In this subsection we will use the expectation value of the energy-momentum tensor component $\mathcal{M}$ and utilize the geometric monodromy technique developed in \cite{Hijano:2018nhq} to obtain a partial expression for the Galilean conformal block in eq. \eqref{4pointmonodromy}. To this end, we obtain the expectation value of the $\mathcal{M}$ component of the energy-momentum tensor from the Ward identities described in eq. \eqref{WardIdentity} as 
\begin{equation}
\mathcal{M}(u_i;(x,t))= \sum_{i=1}^4\left[\frac{h_{M,i}}{(t-t_i)^2}+\frac{C_M}{6}\frac{c_i}{t-t_i}\right],
\end{equation}
where $u_i \equiv (x_i,t_i)$ and the auxiliary parameters $c_i$ are given by
\begin{equation}\label{auxC}
c_i=\frac{6}{C_M} \, \partial_{x_i} \log \left<\sigma_{g_{A}}(u_{1})\sigma_{g_{A}^{-1}}(u_{2})\sigma_{g_{B}}(u_{3})\sigma_{g_{B}^{-1}}(u_{4})\right>\,.
\end{equation}
Here the conformal symmetry does not completely fix the structure of the four point function and hence not all auxiliary parameters $c_i$ are known. Using conformal transformations we place the corresponding twist fields operators at $t_1=0, \,t_3=1, \,t_4=\infty$ leaving $t_2=T$ to be free\footnote{The coordinate $T$ used here is same as the non-relativistic cross ratio $T$ given in \cref{GCFT cross ratio} in terms of the coordinates $t_i$.}. Requiring the scaling of the expectation value $\mathcal{M}(T;t) \sim \,t^{-4}$ as $t \to \infty$ and the fact that in the replica limit $n \to 1$ and $m \to 1$  the conformal dimension $h_{M,i} \equiv h_M$ of the light operator $\sigma_{g_{A}}, \sigma_{g_{B}}$ vanishes, we may express three of the four auxiliary parameters in terms of the fourth. The expectation value of the energy-momentum tensor component $\mathcal{M}$ may then be written as \cite{Basu:2021axf}
\begin{equation}
\frac{6}{C_M}\mathcal{M}(T;t)=c_2\left[\frac{T-1}{t}+\frac{1}{t-T}-\frac{T}{t-1}\right]\,.
\end{equation}
Now under the generic Galilean conformal transformation described in eq. \eqref{finitetransf}, the component $\mathcal{M}$ of the energy-momentum tensor transforms as \cite{Hijano:2018nhq}
\begin{equation}
\mathcal{M}^{\prime}(t^{\prime},x^{\prime})=(f^{\prime})^2\mathcal{M}(t,x)+\frac{C_M}{12}\,S(f,t)\,,
\end{equation}
where $S(f,t)$ is the Schwarzian derivative for the coordinate transformation $t\to f(t)$. For the ground state, the expectation value $\mathcal{M}(u_i;(x,t))$ vanishes on the $GCFT_{1+1}$ plane which leads to the following expression
\begin{equation}\label{Schwarzian}
\frac{1}{2}\,S(f,t)=c_2\left[\frac{T-1}{t}+\frac{1}{t-T}-\frac{T}{t-1}\right]\, .
\end{equation}
This may be expressed in the form of a differential equation as \cite{Basu:2021axf}
\begin{equation}
0= h^{\prime\prime}(t)+\frac{1}{2}S(f,t)\,h(t)=h^{\prime\prime}(t)+\frac{6}{C_M}\mathcal{M}(T,t)\,h(t)\,,
\end{equation}
where the transformation $f$ has the form $h_1/h_2$ with $h_1$ and $h_2$ as the two solutions of the differential equation. This differential equation may be solved by the method of variation of parameters up to linear order in the parameter $\epsilon_{\alpha} = \frac{6}{C_M} h_{M, \alpha}$ which is the rescaled weight of the corresponding conformal block $\mathcal{F}_\alpha$. The monodromy of the solutions whilst circling the light operators at $t=1\,,T$ leads to the following monodromy matrix \cite{Basu:2021axf}:
\begin{equation}
M =\begin{pmatrix}
1 & 2\pi i \,c_2 T(T-1)\\
2\pi i \,c_2(T-1) & 1
\end{pmatrix}\,.
\end{equation}
We may now use the monodromy condition for the three point twist correlator given as \cite{Basu:2021axf}
\begin{equation}\label{THREE}
\sqrt{\frac{I_1-I_2}{2}}=2\pi\epsilon_{\alpha}\,,
\end{equation}
where $I_1=\text{tr}\, M$ and $I_2=\text{tr}\, M^2$ are invariant under global Galilean conformal transformations. Using \cref{THREE}, the auxiliary parameter $c_2$ may now be obtained as follows
\begin{equation} \label{c2}
c_2=\epsilon_{\alpha}\frac{1}{\sqrt{T}(T-1)}\,.
\end{equation}
This provides the form of the conformal block for the four-point function in eq. \eqref{4pointmonodromy} to be
\begin{equation}\label{BLOCK}
\begin{aligned}
\mathcal{F}_{\alpha}&=\exp\left[\frac{C_M}{6}\int\,c_2\,dX\right]\\
&=\exp\left[h_{M,\alpha}\left(\frac{X}{\sqrt{T}(T-1)}\right)\right]\tilde{\mathcal{F}}(T)\,.
\end{aligned}
\end{equation}
Here $\tilde{\mathcal{F}}(T)$ is an unknown function of the coordinate $T$ which will be determined by the monodromy analysis for the other component of the energy-momentum tensor $\mathcal{L}$ in the next subsection.


\subsubsection{Monodromy of $\mathcal{L}$}
In this subsection, we will obtain the complete expression for the Galilean conformal block $\mathcal{F}_\alpha$ in \cref{4pointmonodromy} by performing the geometric monodromy analysis of the energy momentum tensor $\mathcal{L}$. To this end, we begin with the expectation value of $\mathcal{L}$ which may be obtained from the Ward identities in \cref{WardIdentity} to be
\begin{equation}\label{Lexpect1}
\frac{6}{C_M}\mathcal{L}(u_i;(x,t))=\sum_{i=1}^4\bigg[\frac{\delta_i}{(t-t_i)^2}-\frac{1}{t-t_i}d_{i}+\frac{2\epsilon_i(x-x_i)}{(t-t_i)^3}+\frac{x-x_i}{(t-t_i)^2}c_{i}\bigg]\,,
\end{equation}
where $\delta_i = \frac{6}{C_M} \, h_{L,i}$, $\epsilon_i = \frac{6}{C_M} \, h_{M,i}$, the auxiliary parameters $c_i$ are defined in eq. \eqref{auxC} and $d_i$ are given as \cite{Hijano:2018nhq}
\begin{equation}\label{auxiliarydi}
\begin{aligned}
d_i=\frac{6}{C_M} \, \partial_{t_i} \log \left<\sigma_{g_{A}}(u_{1})\sigma_{g_{A}^{-1}}(u_{2})\sigma_{g_{B}}(u_{3})\sigma_{g_{B}^{-1}}(u_{4})\right>\,.
\end{aligned}
\end{equation}
Similar to the previous subsection, by utilizing the global Galilean conformal symmetry we place the twist field operators\footnote{Similar to the case of monodromy analysis of $\mathcal{M}$, the coordinates $X$ and $T$ here are the usual $GCFT$ cross-ratios in terms of the coordinates $(x_i,t_i)$.} at $t_1=0,\, t_2=T,\, t_3=1,\, t_4=\infty$ and $x_1=0,\, x_2=X,\, x_3=0$ and $x_4=0$. Again requiring that $\mathcal{L}$ scales as $\mathcal{L}(T,t)\rightarrow t^{-4}$ with $t\rightarrow \infty$ fixes three of the auxiliary parameters $d_i$ in terms of the remaining one. We may express \cref{Lexpect1} in terms of the undetermined auxiliary parameter $d_2$ as
\begin{equation}
\begin{aligned}
\frac{6}{C_M}\mathcal{L}(u_i;(x,t))=&-\frac{c_2 X+d_2 (T-1)-2 \delta _L}{t}+\frac{c_2 X+d_2 T-2 \delta _L}{t-1}+\frac{c_1 x}{t^2}+\frac{c_2 (x-X)}{(t-T)^2}+\frac{c_3 x}{(t-1)^2}\\
&-\frac{d_2}{t-T}+\frac{2 x \epsilon _L}{t^3}+\frac{\delta
	_L}{t^2}+\frac{\delta _L}{(t-1)^2}+\frac{\delta _L}{(t-T)^2}+\frac{2 \epsilon _L
	(x-X)}{(t-T)^3}+\frac{2 x \epsilon _L}{(t-1)^3}\,.\\
\end{aligned}
\end{equation}
where $\delta_L=\frac{6}{C_M} h_{L}$ and $\epsilon_L=\frac{6}{C_M} h_{M}$ are the rescaled weights of the twist operator $\sigma_{g_{A}}, \sigma_{g_{B}}$. Note that the auxiliary parameters $c_i$s in the above expression are as obtained in the previous subsection \ref{M-monodromy}. On utilizing the transformation of the energy-momentum tensor $\mathcal{L}$ under a generic Galilean transformation \cref{finitetransf}, we arrive at the following differential equation 
\begin{equation}\label{ScaryDE}
\frac{6}{C_M}\mathcal{L}(u_i;(x,t))=\frac{g' \left(f' f''-3 \left(f''\right)^3\right)+f' \left(3 g'' f''-g''' f'\right)}{2\left(f'\right)^3}-\frac{x \left(3 \left(f''\right)^2+f''' \left(f'\right)^2-4 f''' f' f''\right)}{2 \left(f'\right)^3}\,.   
\end{equation}

As described in \cite{Hijano:2018nhq}, we now consider the following combination of the expectation values of the two components of the energy-momentum tensor
\begin{equation}
\mathcal{\tilde{L}} (u_i; (x,t))= \left[ \mathcal{L} (u_i; (x,t)) + X\, \mathcal{M}' (u_i; (x,t)) \right].
\end{equation}
Now we choose $g(t)=f'(t) Y(t)$ as an ansatz for the conformal transformation to reduce the above equation to the following form
\begin{equation}\label{Ltilda}
\frac{6}{C_M}\mathcal{\tilde{L}}=-\frac{1}{2}Y'''-2 Y'\frac{6}{C_{M}}\mathcal{M}-Y\frac{6}{C_M}\mathcal{M'}\,.
\end{equation}
Similar to the $\mathcal{M}$ monodromy in the previous subsection, we may now solve this differential equation up to linear order of the rescaled weights $\epsilon_{\alpha}$ and $\delta_{\alpha}=\frac{6}{C_M} h_{L, \alpha}$ of the corresponding conformal block $\mathcal{F}_\alpha$. Through the monodromy of the solutions circling around the light operators at $t=1\, , T$, we may obtain the undetermined auxiliary parameter $d_2$ as 
\begin{equation}
d_2= \frac{(1-3 T) X \epsilon _{\alpha }+2 (T-1) T \delta _{\alpha }}{2 (T-1)^2 T^{3/2}}\,.
\end{equation}
Finally we may now obtain the full expression for the Galilean conformal block by utilizing \cref{auxiliarydi} to be
\begin{equation}\label{Gconformalblock}
\mathcal{F}_{\alpha} =\left(\frac{1-\sqrt{T}}{1+\sqrt{T}}\right)^{h_{L,\alpha}} \exp\left[h_{M,\alpha}\left(\frac{X}{\sqrt{T }(T-1)}\right)\right]\,.
\end{equation}

We will utilize the above expression for the Galilean conformal block to obtain the reflected entropy for the bipartite mixed state configurations of two disjoint intervals in $GCFT_{1+1}$s in the following subsections.


\subsubsection{Two disjoint intervals at zero temperature}
In this subsection we utilize the large central charge limit expression for the Galilean conformal blocks $\mathcal{F}_{\alpha}$ to obtain the reflected entropy for the bipartite mixed state of two disjoint intervals in a $GCFT_{1+1}$ at zero temperature. In the $t$-channel described by $T\to 1\,,X\to 0$, the dominant contribution to the four-point twist correlator in \cref{4pointmonodromy} comes from the $GCA_2$ conformal block corresponding to the primary field $\sigma_{g_{B}g_{A}^{-1}}$. Also the four point twist correlator in the denominator of \cref{SR-GCFT-def} may be obtained by taking $n \to 1$ limit of \cref{4pointmonodromy} as follows
\begin{equation}\label{deno.}
\begin{aligned}
&\left<\sigma_{g_{m}}(x_{1},t_{1})\sigma_{g_{m}^{-1}}(x_{2},t_{2})\sigma_{g_{m}}(x_{3},t_{3})\sigma_{g_{m}^{-1}}(x_{4},t_{4})\right>_{GCFT^{\otimes m}}\\
& \qquad \qquad \qquad \qquad \qquad \quad =\lim_{n \to1} \left<\sigma_{g_{A}}(x_{1},t_{1})\sigma_{g_{A}^{-1}}(x_{2},t_{2})\sigma_{g_{B}}(x_{3},t_{3})\sigma_{g_{B}^{-1}}(x_{4},t_{4})\right>_{GCFT^{\otimes mn}}\\
&\qquad \qquad \qquad \qquad \qquad \quad=\lim_{n \to1}\left[t_{14}^{-2h_{L}} \, t_{23}^{-2h_{L}} \, \exp\left(-2h_{M} \frac{x_{14}}{t_{14}}-2h_{M} \frac{x_{23}}{t_{23}}\right)\,\sum_{\alpha}\, \mathcal{F}_{\alpha} \left( T, \frac{X}{T} \right)\right], 	
\end{aligned}
\end{equation}
Now by using \cref{4pointmonodromy}, \cref{Gconformalblock} and \cref{deno.}, we may obtain the reflected entropy for two disjoint intervals as follows 
\begin{equation}\label{Dis. result(GCFT)}
S_{R}(A:B)=\frac{C_L}{6} \log \left(\frac{1+\sqrt{T}}{1-\sqrt{T}}\right)+\frac{C_M}{6} \frac{X}{\sqrt{T}(1-T)}+\ldots,
\end{equation}
where $X$ and $T$ are the non-relativistic cross ratios given in eq. \eqref{GCFT cross ratio}. Note that the bulk dual for this case is $(2 + 1)$- dimensional asymptotically flat TMG on a Minkowski space time. We observe that the above expression for the reflected entropy in the large central charge limit matches with twice the corresponding EWCS obtained in \cite{Basu:2021awn}. Here the first term corresponds to twice the CS contribution and the second term involves twice the contribution from the Einstein gravity to the bulk EWCS. Interestingly, taking the appropriate adjacent limit given by $(x_2,t_2) \to (x_3,t_3)$ of our result for the disjoint intervals in \cref{Dis. result(GCFT)}, we reproduce the corresponding adjacent intervals result which constitute a further consistency check for our analysis.


\subsubsection{Two disjoint intervals in a finite size system}		
For this case we consider the two disjoint intervals given by $A \equiv [(\xi_1,\rho_1), (\xi_2,\rho_2)]$ and $B \equiv [(\xi_3,\rho_3), (\xi_4,\rho_4)]$ in a $GCFT_{1+1}$ described on a cylinder of circumference $L$. Similar to the previous subsections, it is necessary to compute the required four point twist correlator on this cylinder. We can map the $GCFT$ complex plane to this cylinder by utilizing \cref{Trans-size-gcft}. Under this conformal map the non-relativistic cross ratios transform as follows
\begin{subequations}\label{modified cross}
	\begin{equation}
	\tilde{T} = \frac{\sin\left(\frac{\pi \xi_{12}}{L}\right)\sin\left(\frac{\pi \xi_{34}}{L}\right)}{\sin\left(\frac{\pi \xi_{13}}{L}\right)\sin\left(\frac{\pi \xi_{24}}{L}\right)}, 
	\end{equation}
	\begin{equation}
	\frac{\tilde{X}}{\tilde{T}} = \frac{\pi \rho_{12}}{L}\cot\left(\frac{\pi \xi_{12}}{L}\right)+\frac{\pi \rho_{34}}{L}\cot\left(\frac{\pi \xi_{34}}{L}\right)-\frac{\pi \rho_{13}}{L}\cot\left(\frac{\pi \xi_{13}}{L}\right)-\frac{\pi \rho_{24}}{L}\cot\left(\frac{\pi \xi_{24}}{L}\right).
	\end{equation}
\end{subequations}
We may obtain the reflected entropy for this bipartite mixed state configuration of two disjoint intervals by simply applying the conformal map \eqref{Trans-size-gcft} in \cref{Dis. result(GCFT)} and utilizing the modified cross ratios in \cref{modified cross} to arrive at
\begin{equation}
S_{R}(A:B)=\frac{C_L}{6} \log \left(\frac{1+\sqrt{\tilde{T}}}{1-\sqrt{\tilde{T}}}\right)+\frac{C_M}{6} \frac{\tilde{X}}{\sqrt{\tilde{T}}(1-\tilde{T})} + \ldots.
\end{equation}
Similar to the previous case, taking the appropriate adjacent limit for the above result reproduces the corresponding expression for the reflected entropy of two adjacent intervals in \cref{adj-L}. As mentioned earlier the bulk dual in this case is described by asymptotically flat TMG on a global
Minkowski orbifold. Note that the above expression for the reflected entropy of the two disjoint intervals in question matches in the large central charge limit with twice the corresponding EWCS computed in \cite{Basu:2021awn}. Here the first term involves twice the CS contribution while the second term
corresponds to twice the global Minkowski orbifold contribution to the bulk EWCS. As earlier these serve as consistency checks for our results.


\subsubsection{Two disjoint intervals at a finite temperature}		
Finally we consider the case of two disjoint intervals  at a finite temperature in a $GCFT_{1+1}$ described on a thermal cylinder with circumference $\beta = 1/T$. Similar to the case of the two adjacent intervals discussed in subsection \ref{sec:adj-T-GCFT}, we employ the conformal transformation given in \eqref{Trans-Temp-gcft} to map the $GCFT$ complex plane to the thermal cylinder. Under this map the $GCFT$ cross ratios are modified as follows
\begin{subequations}\label{Modified}
	\begin{equation}	
	T^\star= \frac{\sinh\left(\frac{\pi \xi_{12}}{\beta}\right)\sinh\left(\frac{\pi \xi_{34}}{\beta}\right)}{\sinh\left(\frac{\pi \xi_{13}}{\beta}\right)\sinh\left(\frac{\pi \xi_{24}}{\beta}\right)},
	\end{equation}
	\begin{equation}  \frac{X^\star}{T^\star} = \frac{\pi \rho_{12}}{\beta} \coth\left(\frac{\pi \xi_{12}}{\beta}\right)+\frac{\pi \rho_{34}}{\beta}\coth\left(\frac{\pi \xi_{34}}{\beta}\right)-\frac{\pi \rho_{13}}{\beta} \coth\left(\frac{\pi \xi_{13}}{\beta}\right)-\frac{\pi \rho_{24}}{\beta}\coth\left(\frac{\pi \xi_{24}}{\beta}\right), 
	\end{equation}
\end{subequations}
where $(\xi, \rho)$ are the coordinates on the cylinder. Now by utilizing \cref{Trans-Temp-gcft} in \cref{Dis. result(GCFT)}, the reflected entropy for the mixed state configuration under consideration may be obtained as follows
\begin{equation}
S_{R}(A:B)= \frac{C_L}{6} \log \left(\frac{1+\sqrt{T^*}}{1-\sqrt{T^*}}\right)+\frac{C_M}{6} \frac{X^*}{\sqrt{T^*}(1-T^*)} + \ldots.
\end{equation}	
Interestingly, the consistency of our results may be checked by implementing the appropriate adjacent limit for the above result which reproduces the corresponding expression for the reflected entropy of two adjacent intervals described in \cref{adj-T}. Note that the holographic dual in this case is described by the bulk non-rotating FSC with TMG. We should also note that in the large central charge limit, the first term of the above expression matches with twice the CS contribution while the second term corresponds to twice the FSC contribution to the bulk EWCS obtained in \cite{Basu:2021awn}.


\section{Summary and conclusion}\label{sec4}

To summarize, in this article we have obtained the reflected entropy for various bipartite pure and mixed state configurations in a class of $(1+1)$-dimensional non-relativistic Galilean conformal field theories. To this end we have established an appropriate replica technique to compute the reflected entropy for bipartite states involving a single, two adjacent and two disjoint intervals in $GCFT_{1+1}$s. In particular, we have computed the reflected entropy for the pure state configuration of a single interval at zero temperature and in a finite sized system and our results exactly match with twice the corresponding entanglement entropies consistent with quantum information theory expectations. Interestingly for the mixed state configuration of a single interval at a finite temperature in a $GCFT_{1+1}$ it was required to employ a construction involving two large but finite auxiliary intervals adjacent to the single interval on either side to compute the reflected entropy in a final bipartite limit. A similar construction has been used in the literature for the computation of the entanglement negativity for this particular mixed state configuration. We also find that our results for the reflected entropy match with twice the upper limit of the bulk EWCS in the dual asymptotically flat geometries computed earlier in the literature in the context of flat holography. This is consistent with the proposed duality between the reflected entropy and the EWCS described earlier in the literature for the usual AdS/CFT scenario.

Subsequent to this, we have obtained the reflected entropy for the mixed state configurations of two adjacent intervals at zero and finite temperatures and in finite sized systems in $GCFT_{1+1}$s through our replica technique. For these case also we observe that our field theory results match with the bulk EWCS for the dual asymptotically flat geometries up to an additive constant arising from the undetermined OPE coefficient of the corresponding three point twist field correlator required to obtain the reflected entropy.

Finally through suitable geometric monodromy analysis of the corresponding four point twist field correlator, we obtain the reflected entropy for the mixed state configuration of two disjoint intervals at zero and finite temperatures and for finite sized systems in the non-relativistic $GCFT_{1+1}$s. We also observe that through appropriate limits of the mixed state configurations involving the two disjoint intervals, we may obtain the results for the configuration of two adjacent intervals and these are consistent with this limiting procedure which constitutes an additional consistency check for our analysis. Additionally, in the appendix \ref{appendix_A} we have also reproduced the expression for the reflected entropy for two disjoint intervals in a $GCFT_{1+1}$ through a parametric contraction of the corresponding result in the usual relativistic $CFT_{1+1}$ which provides a strong substantiation for our computations. Furthermore we mention here that for all the three cases involving the two disjoint intervals, our results for the reflected entropy match exactly with twice the bulk EWCS obtained earlier in the literature in the context of flat space holography. Through these results we conclude that the holographic duality between the reflected entropy and the EWCS, {\it i.e.,} $S_R(A:B) = 2 E_W(A:B)$, also holds in flat space holographic scenarios involving bulk asymptotically flat $(2+1)$-dimensional geometries dual to $GCFT_{1+1}$s.
\bigskip

\noindent {\bf Note added:} While this article was in its final stage of communication \cite{Setare:2022uid} appeared in the e-print arXiv which obtained some of the results discussed in this article in a somewhat different context.


\appendices
\section{Limiting Analysis} \label{appendix_A}
In this appendix we will show through an example that the reflected entropy for bipartite mixed states in $GCFT_{1+1}$s computed in \cref{sec3} may be obtained through a parametric contraction of the corresponding result obtained in context of the relativistic $CFT_{1+1}$s in \cite{Dutta:2019gen, Jeong:2019xdr}. To this end, the $GCA_{1+1}$ algebra may be obtained through a parametric \.In\"on\"u-Wigner contraction given in \cref{Inonu-Wigner} of the usual Virasoro algebra for relativistic $CFT_{1+1}$s. The \.In\"on\"u-Wigner contraction may alternatively be written in terms of the coordinates describing the $CFT_{1+1}$ complex plane as
\begin{equation}\label{Contractions}
z \to t + \epsilon x \, \, , \qquad \bar{z} \to t - \epsilon x \, ,
\end{equation}
with $\epsilon \to 0$. We may also relate the central charges of the $GCA_{1+1}$ to those of the parent relativistic theory as \cite{Basu:2021awn}
\begin{equation}\label{gcabms}
C_L = c + \bar{c} \quad, \quad C_M = \epsilon(c - \bar{c}) \, .
\end{equation}

We will now utilize the above to illustrate that the results for the reflected entropy in $GCFT_{1+1}$s are consistent with those in usual $CFT_{1+1}$s under the above non-relativistic limit. To this end, we recall the expression for the reflected entropy for the generic bipartite mixed state of two disjoint intervals at zero temperature in a $CFT_{1+1}$ in the $t$-channel to be \cite{Dutta:2019gen, Jeong:2019xdr}
\begin{equation}
S_R (A : B) = \frac{c}{3} \log \left( \frac{1 + \sqrt{x}}{1 - \sqrt{x}} \right) 
\end{equation}
where $x=\frac{z_{12}z_{34}}{z_{13}z_{24}}$ is the $CFT_{1+1}$ cross ratio. If we allow unequal central charges for the holomorphic and anti-holomorphic sectors, the above expression may be written as
\begin{equation}\label{SR-CFT}
S_R (A : B) = \frac{c}{6} \log \left( \frac{1 + \sqrt{x}}{1 - \sqrt{x}} \right) + \frac{\bar{c}}{6} \log \left( \frac{1 + \sqrt{\bar{x}}}{1 - \sqrt{\bar{x}}} \right) .
\end{equation}
Utilizing \cref{Contractions}, we may now express the $CFT_{1+1}$ cross ratios $x, \bar x$ in terms of the $GCFT_{1+1}$ cross ratios $X, T$ as
\begin{equation}\label{CrossTrans}
x\rightarrow T\left(1+\epsilon \frac{X}{T}\right)\,\,,\qquad \bar{x}\rightarrow T\left(1-\epsilon \frac{X}{T}\right)\,.
\end{equation}
Using the above, the reflected entropy in $GCFT_{1+1}$ may be obtained through the \.In\"on\"u-Wigner contraction of \cref{SR-CFT} up to linear order in $\epsilon$ to be
\begin{equation}
S_R (A : B) = \frac{C_L}{6} \log \left( \frac{1 + \sqrt{T}}{1 - \sqrt{T}}\right) + \frac{C_M}{6} \frac{X}{\sqrt{T} (1 - T)} + \mathcal{O} (\epsilon) .
\end{equation}
Remarkably, in the leading order the above expression matches exactly with the replica technique result obtained in \cref{Dis. result(GCFT)} which provides a strong consistency check for our computations. We have also checked that the reflected entropy for the other bipartite states in $GCFT_{1+1}$s discussed in this article are also consistent with this limiting behaviour. Although, we may obtain the results for the reflected entropy for subsystems in $GCFT_{1+1}$s through the above limiting analysis, however, note that it does not provide any information about the structure of the replicated manifold which is necessary in the study of the R\'enyi reflected entropy and its applications in different contexts including holography.


	\bibliographystyle{utphys}
	
	\bibliography{reference}
	
\end{document}